\documentclass[a4paper,11pt]{article}

\pdfoutput=1

\usepackage{jinstpub}
\usepackage{amsmath}
\usepackage{siunitx} 
\usepackage[section]{placeins} 
\usepackage{upgreek} 
\usepackage[shortcuts]{extdash} 
\usepackage{booktabs} %
\usepackage{multirow} %
\usepackage{makecell} 

\graphicspath{{figures/}}

\newcommand{\fig}[1]{figure~\ref{fig:#1}} %
\newcommand{\Fig}[1]{Figure~\ref{fig:#1}} 

\newcommand{\eq}[1]{equation~\ref{eq:#1}} %
\newcommand{\Eq}[1]{Equation~\ref{eq:#1}} 

\newcommand{\Tab}[1]{Table~\ref{tab:#1}} 

\newcommand{\sect}[1]{section~\ref{sec:#1}} %

\newcommand{\samplemean}[1]{\overline{#1}} 
\newcommand{\var}[1]{\mathrm{var}\!\left({#1}\right)} 
\newcommand{\cov}[2]{\mathrm{cov}\!\left({#1,\,#2}\right)} 
\newcommand{\vect}[1]{\mathbf{#1}} 
\newcommand{\mat}[1]{\mathbf{#1}} 
\newcommand{\transpose}[1]{{#1}^{\intercal}} 

\DeclareSIUnit\electron{\ensuremath{\textnormal{e}^-}} 
\DeclareSIUnit\clight{\text{\ensuremath{\mathit{c}}}} 
\DeclareSIUnit\evperc{\eV/\clight} 

\title{Timing performance of the {LHCb}~{VELO}~{Timepix3}~Telescope}

\author[a,1]{K.~Heijhoff,\note{Corresponding author.}}
\author[a]{K.~Akiba,}
\author[a]{M.~van~Beuzekom,}
\author[a]{P.~Bosch,}
\author[b]{J.~Buytaert,}
\author[b]{M.~Campbell,}
\author[a]{A.P.~Colijn,}
\author[b]{P.~Collins,}
\author[a,2]{E.~Dall'Occo,\note{Now at Fakult\"at Physik, Technische Universit\"at Dortmund, Dortmund, Germany}}
\author[c,3]{T.~Evans,\note{Now at CERN, Geneva, Switzerland}}
\author[a]{R.~Geertsema,}
\author[a]{M.L.E.~Heidotting,}
\author[a]{D.~Hynds,}
\author[b]{X.~Llopart Cudie,}
\author[b]{H.~Schindler,}
\author[a]{H.~Snoek}

\affiliation[a]{Nikhef, Amsterdam, the Netherlands}
\affiliation[b]{CERN, Geneva, Switzerland}
\affiliation[c]{Department of Physics, University of Oxford, Oxford, United Kingdom}

\emailAdd{k.heijhoff@nikhef.nl}

\abstract{We performed a detailed study of the timing performance of the LHCb VELO Timepix3 Telescope with a \SI{180}{\giga\evperc} mixed hadron beam at the CERN SPS. A twofold method was developed to improve the resolution of single\=/plane time measurements, resulting in a more precise overall track time measurement. The first step uses spatial information of reconstructed tracks in combination with the measured signal charge in the sensor to correct for a mixture of different effects: variations in charge carrier drift time; variations in signal induction, which are the result of a non\=/uniform weighting field in the pixels; and lastly, timewalk in the analog front\=/end. The second step corrects for systematic timing offsets in Timepix3 that vary from \SIrange{-2}{2}{\nano\second}. By applying this method, we improved the track time resolution from \SI{438(16)}{\pico\second} to \SI{276(4)}{\pico\second}.}


\begin{document}
	\maketitle
	\flushbottom
	
	\section{Introduction}
	The LHCb VELO Timepix3 Telescope \cite{Akiba:2019} was constructed to characterise sensor prototypes for the upgrade of the LHCb Vertex Locator \cite{LHCbVELOGroup:2014}. The telescope planes use the Timepix3 pixel ASIC \cite{Poikela:2014}, which provides simultaneous measurements of both time and charge for all hits. The single\=/hit time resolution is limited to at least \SI{451}{\pico\second} by the TDC bin size of \SI{1.56}{\nano\second}. In practice the time resolution is degraded further due to jitter in the analog front\=/end and variations in the TDC bin size.
	
	In view of the High Luminosity LHC \cite{Apollinari:2017}, it is foreseen that precise time measurements will become crucial in the track reconstruction of particle physics experiments. As the number of events per bunch crossing (pile\=/up) will increase, it is expected that experiments will need to use 4D~tracking \cite{Cartiglia:2017, Sadrozinski:2018} in which time measurements are used in the track reconstruction. The additional temporal information on tracks enables the reconstruction algorithm to better distinguish spatially overlapping vertices, and thus increase the maximum pile\=/up that the experiment can cope with. To study sensor prototypes for a future 4D tracker, a fast\=/timing telescope will be constructed using the Timepix4 ASIC \cite{Ballabriga:2020}, which is the successor of Timepix3 and has a TDC bin size of \SI{195}{\pico\second}. 
	
	This paper describes a detailed study of the timing performance of the current Timepix3\=/based telescope with a focus on both the sensor as well as the ASIC. Currently, the uncertainty in the time measurement is still dominated by the sensor, but as the community moves to faster sensor technologies, the ASIC will become more important, and could become the limiting factor. Understanding the timing characteristics of the ASIC will therefore be essential in taking full advantage of fast sensor technologies. Additionally, this study also addresses the problem of combining the time measurements of different detector planes into an overall track time measurement. Since the track reconstruction of the Timepix3 telescope relies heavily on time measurements, it is in principle a 4D~tracker, and the future Timepix4~telescope will take the timing performance to the next level. The methodology that is used in this study to optimise the overall track time resolution may also be applied in a future 4D tracker.

	\section{Experimental setup}
	\subsection{The LHCb VELO Timepix3 Telescope}
\begin{figure}[htbp]
		\centering
		\includegraphics{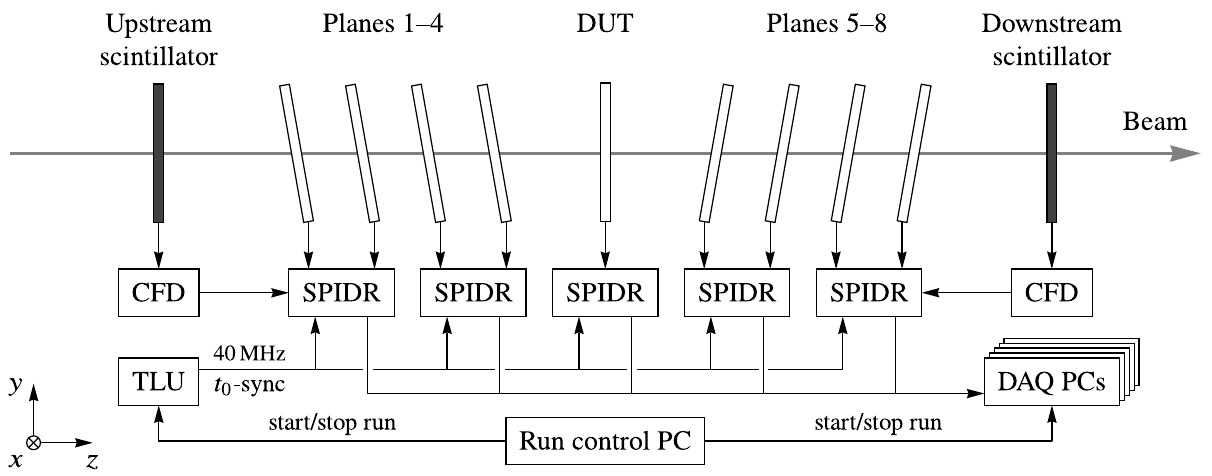}%
		\caption{Diagram of the LHCb VELO Timepix3 Telescope.}
		\label{fig:telescopeDiagram}
	\end{figure}

	\Fig{telescopeDiagram} shows a diagram of the Timepix3 telescope and defines the reference frame. It has eight planes that consist of p\=/on\=/n silicon sensors with a thickness of \SI{300}{\micro\meter} and ${55\times\SI{55}{\square\micro\meter}}$ pixels. The sensors are bonded to Timepix3 ASICs, which provide simultaneous measurements of time and charge for each hit. The telescope planes are rotated by approximately~\SI{9}{\degree} around both the $x$- and $y$\=/axis to introduce charge sharing and hence optimise the spatial resolution. As a result, \SI{59}{\percent} of the clusters consist of 3 hits, and less than \SI{1}{\percent} only have a single hit.
	
	During typical operation the sensors are biased at \SI{150}{\volt} in order to optimise the charge sharing for spatial resolution, but in this study we biased them at \SI{200}{\volt} to optimise the timing performance by increasing the drift velocity of charge carriers. The bias voltage is limited to this value due to the risk of breakdown as the smallest distance between HV and GND is only \SI{30}{\micro\meter}. Our measurements were performed using a beam of mixed hadrons ($p$, $\pi$, $K$) of about \SI{180}{\giga\evperc} at the H8 beam line of the CERN Super Proton Synchrotron (SPS). The hadrons were delivered in spills that are repeated every \SIrange[range-phrase=--, range-units=single]{20}{30}{\second} and contain a few million particles that are distributed over a duration of typically~\SI{4.5}{\second}.
	
	\subsection{Track reconstruction}
	We use the Gaudi based software framework called \textit{Kepler} \cite{Akiba:2019} to spatially align the telescope and reconstruct the tracks. The reconstruction algorithm first aligns the telescope planes in time and makes clusters from neighbouring hits that fall within a time window of \SI{100}{\nano\second}. The charge\=/weighted mean of the pixels in a cluster determine the cluster position in each plane. The algorithm finds clusters that belong to the same track and performs straight\=/line fits in the $yz$- and $zx$\=/planes. We use the standard track reconstruction conditions as discussed in \cite{Akiba:2019} except for a tighter $\chi_\nu^2$ cut of the track fit, which we reduced from $\chi_\nu^2<10$ to $\chi_\nu^2<5$ to improve the track quality. Important to reiterate here is that the track reconstruction requires a single cluster on each of the eight telescope planes, which means that the track time is always calculated using eight planes.
	
\begin{figure}[htbp]
		\centering
		\includegraphics{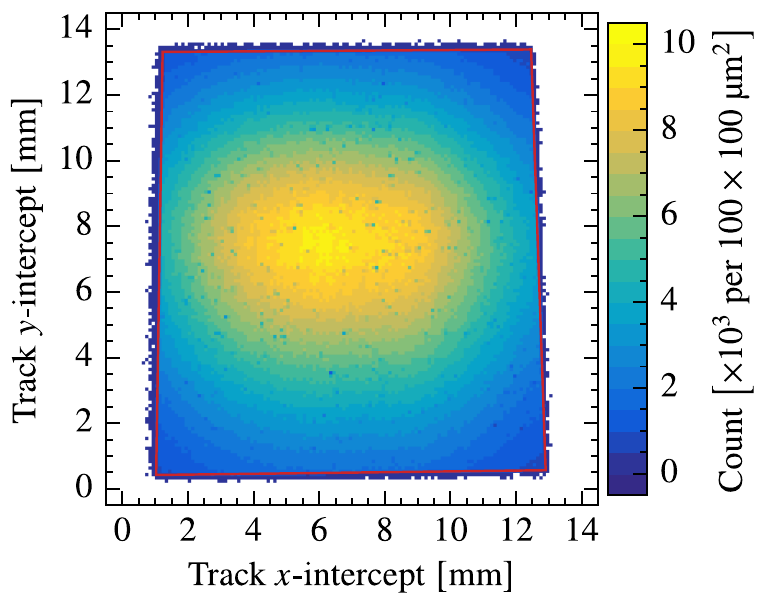}%
		\caption{Map of the reconstructed track positions at the $z$\=/centre of the telescope showing the beam profile. The red outline indicates the intersection of the telescope planes in the beam direction}
		\label{fig:beamHeatMap}
	\end{figure}

	The data sample that is used for this study consists of \num{72.2e6} reconstructed tracks out of which we reserve \num{71.2e6} tracks to determine the corrections that are discussed in the subsequent sections. We use the remaining \num{1e6} tracks to determine the time resolution before and after applying the corrections. \Fig{beamHeatMap} shows the beam profile in the $xy$\=/plane (perpendicular to the beam direction) as measured at the z\=/centre of the telescope where the best track pointing resolution is achieved. The beam has a FWHM of about \SI{11}{\milli\metre} in the $x$\=/direction and \SI{8.2}{\milli\metre} in the $y$\=/direction. Masked pixels appear as shadows and form spots of lower intensity. The red outline highlights the intersection between the telescope planes in the beam direction. A small fraction of the tracks (\num{4.7e-4}) falls outside of this region due to deviations in the track direction.
	
	\subsection{Scintillator time measurement using the SPIDR TDC}
	We used fast scintillators with an active area of ${1.5\times\SI{1.5}{\square\centi\meter}}$ to provide two independent time references. They are located up- and downstream of the telescope (as was shown in \fig{telescopeDiagram}), and are equipped with constant fraction discriminators (CFD). As will be described in \sect{rawTimingPerformance}, we determined their time resolutions to be \SI{381(8)}{\pico\second} and \SI{182(4)}{\pico\second}, respectively. Unless stated otherwise, we use the downstream scintillator as a reference for the time residuals.
	
\begin{figure}[htbp]
		\centering
		\includegraphics{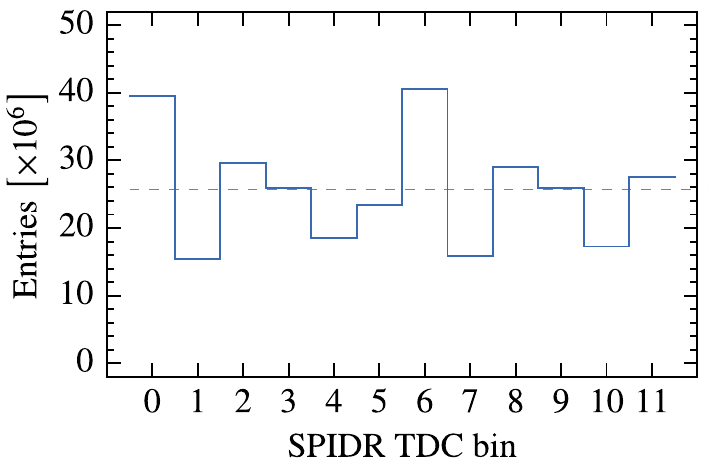}%
		\caption{Histogram of the SPIDR TDC time bins. The dashed line shows the expected distribution for time bins of equal size.}
		\label{fig:spidrTdcFineTimeHistogram}
	\end{figure}

	The on\=/board TDC of the SPIDR readout \cite{Visser:2015, Heijden:2017} provides the time measurement of the scintillator signals. The TDC is achieved by six phase\=/shifted \SI{320}{\mega\hertz} clocks dividing the \SI{320}{\mega\hertz} clock period into twelve time bins of~\SI{260}{\pico\second}. \Fig{spidrTdcFineTimeHistogram} shows that the time bins vary slightly in size as a result of variations in the phase shifts of the six clocks. To correct for this effect, we assume that the particles are uncorrelated to the clock, and measure the occupancy of each bin to determine its actual size.
	
\begin{figure}[htbp]
		\centering
		\includegraphics{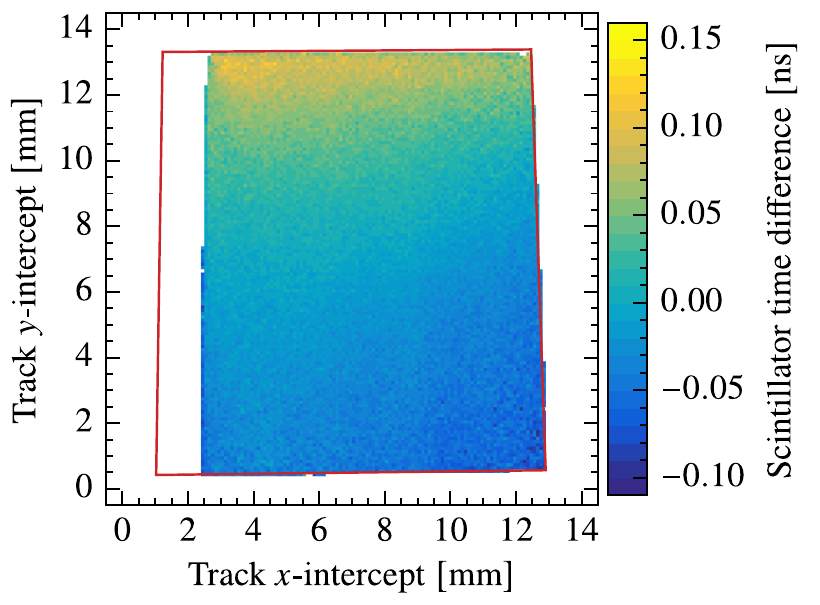}%
		\hskip10mm
		\includegraphics{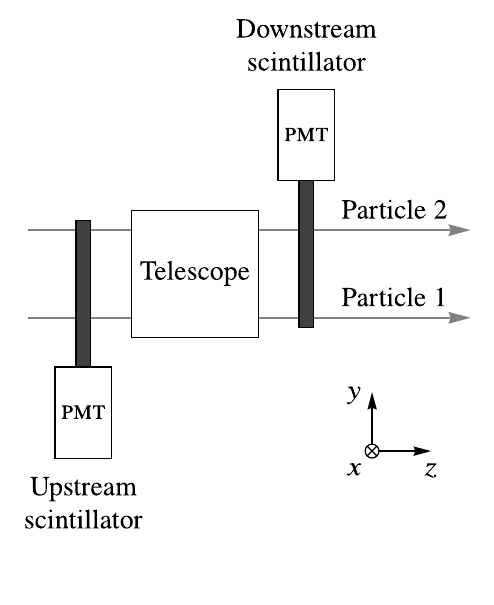}%
		\caption{Mean time\=/difference between the up- and downstream scintillators as a function of the transverse track position (left) and a diagram showing the scintillator placement (right). The red outline indicates the overlapping region of the telescope planes. The left region of the plot is outside of the acceptance region of the downstream scintillator.}
		\label{fig:scintTimeDiffVsTrackPos}
	\end{figure}

	\Fig{scintTimeDiffVsTrackPos} shows the mean time\=/difference between the up- an downstream scintillators, which is calculated as $\samplemean{\Delta t} \equiv \samplemean{t_{\textnormal{u}}-t_{\textnormal{d}}}$. It can be seen that there is a clear spatial dependence: at low $y$\=/positions the upstream scintillator is too early with respect to the downstream scintillator, and it is too late at high $y$\=/positions. This can be explained by considering the two particles in the diagram of \fig{scintTimeDiffVsTrackPos}. For particle one it can be seen that the transit time\===the time it takes for scintillation light to travel to the PMT\===will be shorter in the upstream scintillator than in the downstream scintillator, and vice versa for particle two. Consequently, the time difference for particle one will be lower than for particle two. The mean time\=/difference between the scintillators at $y=\SI{1}{\milli\meter}$ and $\SI{13}{\milli\meter}$ differs by about $\SI{0.13}{\ns}$. From this we estimate the average transit speed in the scintillators to be

	\begin{equation*}
		v = 2 \times \frac{\SI{12}{\milli\meter}}{\SI{0.13}{\nano\second}} = \SI{18}{\centi\meter\per\second}
		\,
	\end{equation*}
	where the factor of two can be understood by supposing that $dt_{\textnormal{u}}/dy=1/v$ and $dt_{\textnormal{d}}/dy=-1/v$, and therefore $d(t_{\textnormal{u}}-t_{\textnormal{d}})/dy = 2/v$. The value we find is similar to the speed of light in a typical plastic scintillator ($n=\num{1.58}$) of \SI{19}{\cm\per\ns}, and the difference can be explained by the fact that the actual transit time is increased by light reflections at the scintillator surfaces \cite{Knoll:2010}.
	
	The spatial dependence of the time difference between the scintillators would introduce a correlation between the scintillator measurements, but we prevent this by correcting the time measurement of the upstream scintillator by subtracting the mean time\=/difference for the current track position: $t_{\textnormal{up}} \rightarrow t_{\mathrm{up}} - \samplemean{\Delta t}\!\left(x,\,y\right)$. This eliminates the spatial dependence of the time difference between the scintillators.

	\subsection{Time measurement with Timepix3} \label{sec:timeMeasurementInTpx3}
\begin{figure}[htbp]
		\centering
		\includegraphics{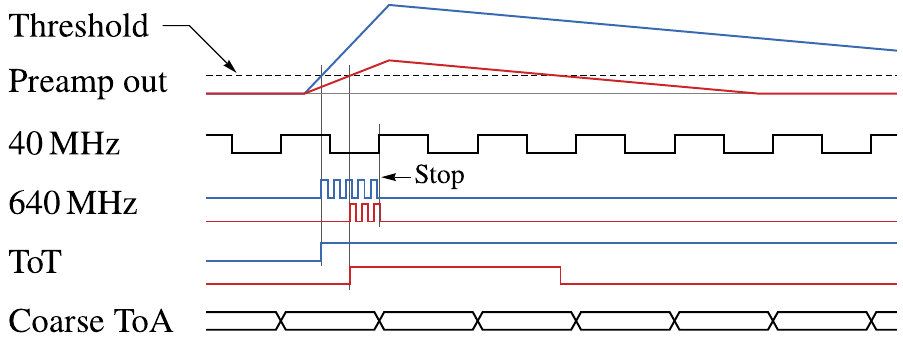}%
		\caption{Illustration of the ToA and ToT measurements in Timepix3.}
		\label{fig:tpx3TimeDiagram}
	\end{figure}

	\Fig{tpx3TimeDiagram} illustrates the time measurement mechanism in Timepix3. A hit is registered when the preamplifier output crosses a threshold that is equivalent to a signal charge of \SI{1000}{\electron}. This starts a \SI{640}{\mega\hertz} oscillator, which runs until the next rising edge of the \SI{40}{\mega\hertz} reference clock. By counting the number of clock cycles of both the oscillator and the reference clock, the time of arrival (ToA) is measured with a granularity of~\SI{1.56}{\nano\second}. Timepix3 also measures the duration that the preamplifier output is above the threshold with a granularity of \SI{25}{\nano\second}. This is known as time over threshold (ToT), and it is a measure for the amount of charge that is collected by the preamplifier. The blue and red signals in the diagram represent two events with different charge. Both are generated at the same time, but the event that has lower charge is measured to be later than the other because of the rise time of the preamplifier. Moreover, the reaction speed of the discriminator is proportional to the time derivative of the signal, $dV/dt$, at the threshold level due to capacitive loading. This dependence of the time measurement on the signal amplitude is known as \textit{timewalk}. It degrades the time resolution, but it can be corrected using the ToT measurement.
	
\begin{figure}[htbp]
		\centering
		\includegraphics{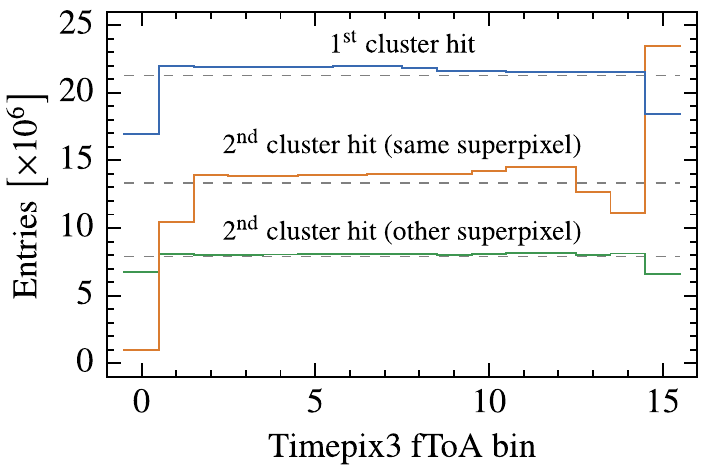}
		\caption{Distributions of the fToA bins for (1) first hits in a cluster, (2) second hits that arrive in the same superpixel as the first hit, and (3) second hits that arrive in a different superpixel than the first. The dashed lines show the mean number of hits per fToA bin.}
		\label{fig:ftoaDistributions}
	\end{figure}

	Imperfections in the fast oscillators affect the uniformity of the time bins, and therefore degrade the time resolution. For instance, not all clock cycles have the same duration. This can be seen by looking at the number of counts from the fast oscillator (\SI{640}{\mega\hertz}), which we refer to as the fine time of arrival (fToA). Its value can range from \num{0} to \num{15} because the fast oscillator divides the \SI{40}{\mega\hertz} period into \num{16} time bins. \Fig{ftoaDistributions} shows histograms of the fToA from first and second hits in a cluster. For the first hits it can be seen that the time bins with an fToA of \num{0} and \num{15} contain less events, which means that, on average, these time bins are smaller than the others. Each fast oscillator in Timepix3 is shared by a group of \num{2} by \num{4} pixels, called a superpixel. We can see that the time bins change when the second hit arrives in the same superpixel as the first. The first time bin becomes smaller because there is no start\=/up time of the oscillator since it is already running, and the last time bin becomes larger because the pixel stops counting clock edges when it reaches~\num{15}. This shows that there is a clear difference in the time measurement when the superpixel's oscillator is already running. The time bins only change slightly when the second hit is in a different superpixel because its fast oscillator is still suspended.
	
	\subsection{Charge calibration}
	\label{sec:chargeCalibration}
	The analog front\=/end of Timepix3 uses a charge sensitive preamplifier that follows the Krummenacher scheme \cite{Krummenacher:1991}. When a transient signal current arrives, it is integrated onto a feedback capacitor, and the feedback circuitry subsequently starts discharging the feedback capacitor with a constant current known as the Krummenacher current. The preamplifier output is directly proportional to the amount of charge on the feedback capacitor, and since the discharge current is constant, the time over threshold will be roughly proportional to the total amount of charge in the signal.

\begin{figure}[htbp]
		\centering 
		\includegraphics{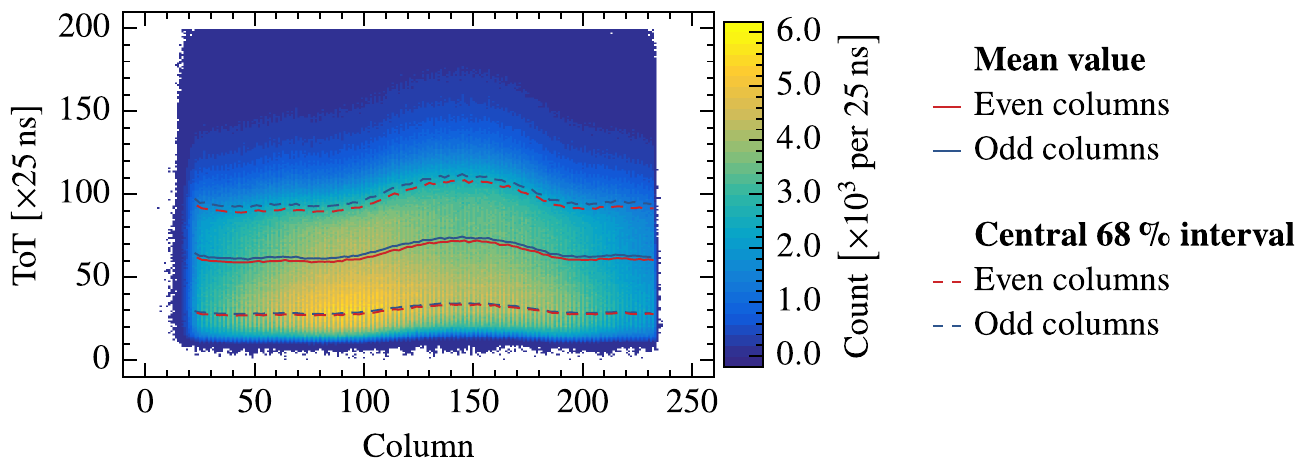}%
		\caption{Time over threshold distribution of first hits in a cluster for all columns of a single telescope plane.}
		\label{fig:totVsCol}
	\end{figure}

	For this study we are interested in the ToT measurement mainly for timewalk corrections. The timewalk correction will be realised by averaging over multiple pixels. However, this degrades the accuracy of the correction because the relationship between ToT and signal charge is not the same for all pixels. This can be seen in \fig{totVsCol}, which shows that the ToT~distribution varies significantly over the columns. This variation cannot originate in the underlying charge distribution because this distribution is expected to be constant as we have a uniform sensor thickness. The prolonged ToT that we observe around column 150 is due to a decreased Krummenacher current\===it takes longer to discharge the same amount of charge.\footnote{The telescope is built with the first iteration of Timepix3. The Krummenacher current uniformity was improved in the second iteration.} Since timewalk depends mainly on charge, the relationship between the ToT and the delay from timewalk is also different among pixels, and consequently we cannot simply use the ToT~measurement directly.

\begin{figure}[htbp]
		\centering
		\includegraphics{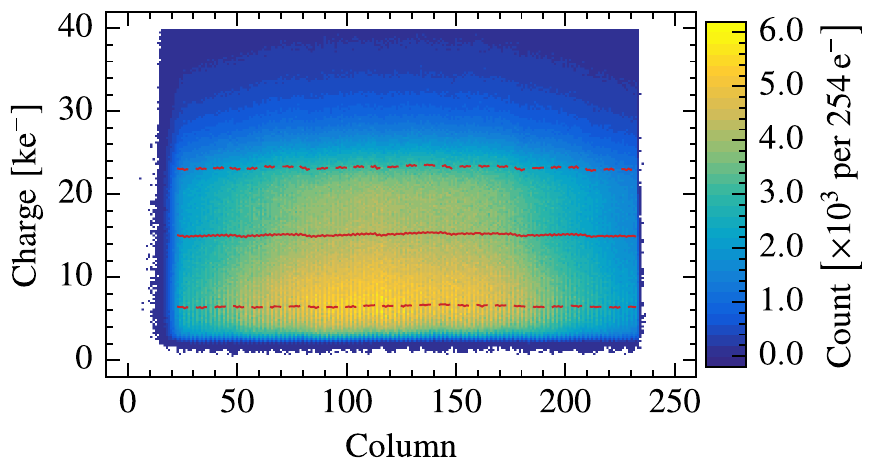}%
		\caption{Charge distribution of first hits in a cluster for all columns of a single telescope plane. The solid curve shows the mean charge, and the dashed curves show the central \SI{68}{\percent} interval.}
		\label{fig:chargeVsCol}
	\end{figure}

	We apply a calibration to each pixel to determine the amount of signal charge from a ToT~measurement. We determine the relationship between charge and ToT for each pixel by injecting a controlled amount of charge into the preamplifier and measuring the ToT response. This relationship is usually modelled as \cite{Jakubek:2008}
	\begin{equation*}
		T\!oT = p_0 + p_1 Q - \frac{p_2}{Q - p_3}
		\,.
	\end{equation*}
	After obtaining the fit parameters $p_{0}$\==$p_{3}$ for each pixel, we can use the inverse relationship to determine the charge from a measured ToT value. \Fig{chargeVsCol} shows the result of the charge calibration. The even\=/numbered columns contain more hits because, as will be shown in \sect{pixelMatrixCorrections}, the time measurement in an even\=/numbered column is, on average, earlier than in an odd\=/numbered column, and since we are only looking at the first hits in a cluster, there is a slight bias towards even\=/numbered columns.
		
\begin{figure}[htbp]
		\centering
		\includegraphics{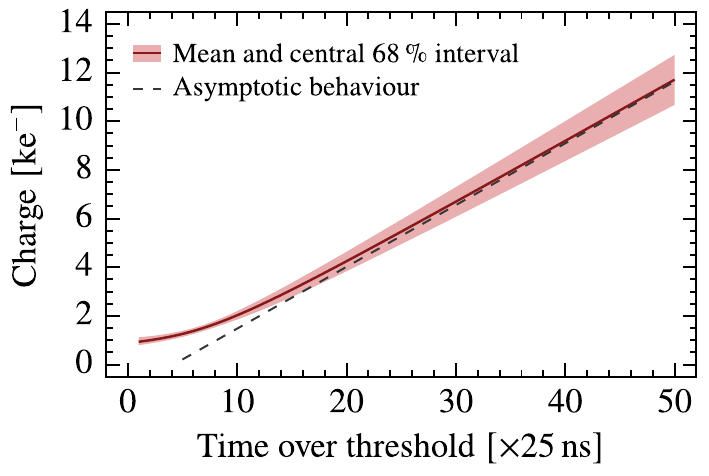}%
		\caption{Charge distribution over the pixels in a single telescope plane as a function of the time over threshold. The solid line shows the mean  of the charge distribution over the pixels, the shaded band shows the central \SI{68}{\percent} interval, and the dashed line is the asymptotic behaviour.}
		\label{fig:chargeCalibration}
	\end{figure}

	\Fig{chargeCalibration} shows the spread in the relationship between charge and ToT for one of the telescope planes. The mean asymptotic behaviour given by 
	\begin{equation*}
		Q = T\!oT \; \samplemean{1/p_1} - \samplemean{p_0/p_1}
	\end{equation*}
	is also shown. Here $\samplemean{1/p_1}$ and $\samplemean{p_0/p_1}$ denote the mean values of $1/p_1$ and $p_0/p_1$ over all pixels. For a sufficiently large ToT, the mean value of $1/p_1$ corresponds to the mean change in charge of a single ToT count difference, and therefore we will use the value of $\samplemean{1/p_1}$ as the charge bin size in the subsequent sections. Its value ranges from about \SIrange{240}{320}{\electron} depending on the telescope plane.
	
	\section{Timing characteristics of the Timepix3 telescope}
	
	\subsection{Raw timing performance} \label{sec:rawTimingPerformance}
\begin{figure}[htbp]
		\centering
		\includegraphics{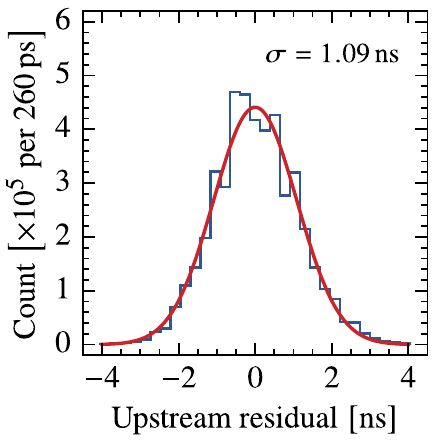}%
		\includegraphics{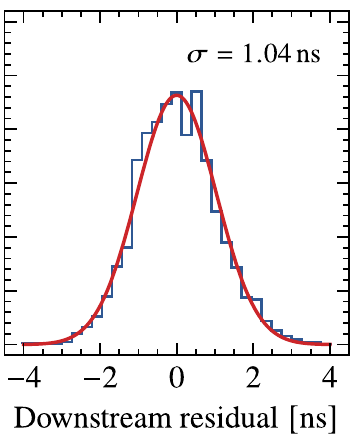}%
		\hskip 5mm%
		\includegraphics{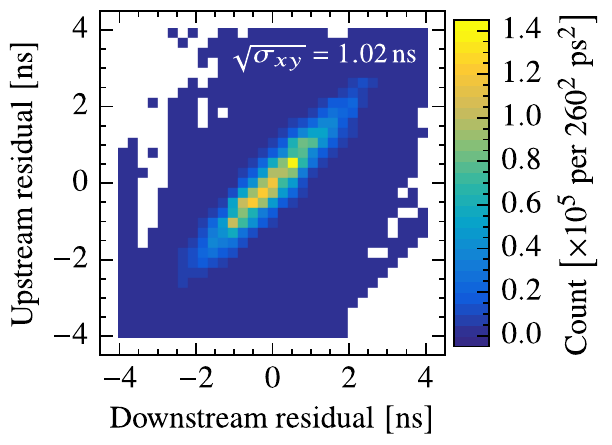}%
		\caption{The two left plots show the time residuals w.r.t. the up- and downstream scintillators of the first hits in a cluster. The right plot shows the correlation between the two residuals. The time resolution can be calculated as the square root of the covariance of the right plot.}
		\label{fig:timeResidualsBefore} 
	\end{figure}

	\Fig{timeResidualsBefore} shows the time residuals of the first hits in a cluster with respect to the up- and downstream scintillators. Since the scintillators provide independent measurements, we can find the time resolution by determining the covariance of the two residuals:
	\begin{equation}
		\label{eq:timeResolution}
		\sigma_{t}^2=\cov{t - t_{\textnormal{d}}}{t - t_{\textnormal{u}}}
	\end{equation}
	where $t$ is the hit time, and $t_{\textnormal{u}}$ and $t_{\textnormal{d}}$ are the up- and downstream scintillator measurements. This gives the time resolution without contributions from the scintillators. The scintillator resolution can be determined as $\sigma^2_{\textnormal{u/d}}=\var{t-t_{\textnormal{u/d}}}-\sigma_{t}^2$. The correlation between the two residuals is clearly visible in \fig{timeResidualsBefore}, which means that the scintillators have sufficient precision to determine the time resolution of the telescope hits.  
	
	\Tab{planeResolutionsRaw} shows the overall cluster time resolution for the first, second, and third hits in a cluster for both oscillator states. The oscillator is always off when the first arrives, but for subsequent hits the oscillator is only off when the hit arrives in a different superpixel than the preceding hits in the cluster. For the first hits, we find an overall resolution of~\SI{1.02(4)}{\nano\second}. The second and third hits are in the range of \SIrange{3.5}{7}{\nano\second}, depending on the telescope plane. The time resolution of these hits is significantly worse because the majority of them have less charge and hence suffer more from timewalk as well as the associated jitter. Correcting for timewalk in these hits only results in a minor improvement. Furthermore, an already running oscillator also leads to a worse time resolution because of changes in the time bin sizes as shown in \sect{timeMeasurementInTpx3}. Because of the disparity in quality between hits, we define the cluster time as that of the first hit.
	
	\begin{table}[htbp]
		\centering
		\caption{Time resolution of the telescope planes for different hits in a cluster and the two possible states of the superpixel oscillator. These values include a \SI{451}{\pico\second} contribution from time binning in Timepix3.}
		\label{tab:planeResolutionsRaw}
		\smallskip
		\small
		\sisetup{table-align-uncertainty=true}
		\begin{tabular}{cS[table-format=1.2(1)]S[table-format=1.1(1)]S[table-format=1.1(1)]}
			\toprule
			           & \multicolumn{3}{c}{Time resolution [\si{\nano\second}]}     \\ \cmidrule{2-4}
			Oscillator & {1\textsuperscript{st} hit} & {2\textsuperscript{nd} hit} & {3\textsuperscript{rd} hit} \\ \midrule 
			off        & 1.02(4) & 4.5(4) & 5.7(3) \\
			on         & {n.a.}  & 5.6(7) & 6.3(4) \\
			\bottomrule
		\end{tabular}
	\end{table}
	
\begin{figure}[htbp]
		\centering
		\includegraphics{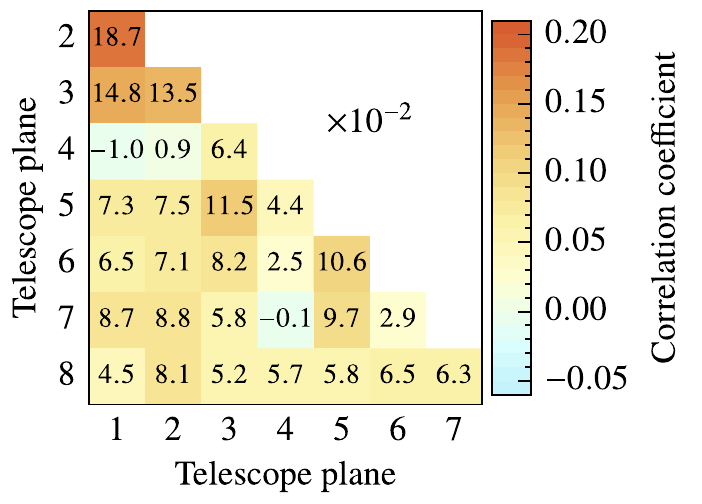}%
		\caption{Correlations between the telescope planes}
		\label{fig:correlationCoefficientsBefore}
	\end{figure}

	The correlation coefficient of time measurements from different planes is given by
	\begin{equation*}
		\rho_{ab}=\frac{\cov{t_a - t_{\textnormal{d}}}{t_b - t_{\textnormal{u}}}}{\sigma_a \sigma_b}
	\end{equation*}
	where $t_{a}$ and $t_{b}$ are the cluster time measurements from telescope planes $a$ and $b$. The time resolutions $\sigma_a$ and $\sigma_b$ are determined by \eq{timeResolution}. We observe a clear correlation between the planes with coefficients ranging from about~\numrange{-0.01}{0.19} as can be seen in \fig{correlationCoefficientsBefore}.
	
	The track time can be determined as a weighted mean that takes the correlations into account by writing down a $\chi^2$ for fitting a single parameter: \cite{Brandt:2014}
	\begin{equation*}
		\chi^2 = \transpose{\left(\vect{t}_{\textnormal{cl}} - \vect{1} \, t_{\textnormal{tk}}\right)} \mat{C}^{-1} \left(\vect{t}_{\textnormal{cl}} - \vect{1} \, t_{\textnormal{tk}}\right)
	\end{equation*}
	where~$t_{\textnormal{tk}}$ is the track time, $\vect{t}_{\mathrm{cl}}=\left(t_1,\,\ldots,\,t_8\right)$ are the cluster time measurements from the telescope planes, $\vect{1}$ is a vector of ones, and $\mat{C}$ is the covariance matrix. Minimising gives a track time calculated as
	\begin{equation}
		\label{eq:trackTime}
		t_{\textnormal{tk}} = \left(\transpose{\vect{1}}\mat{C}^{-1}\vect{1}\right)^{-1} \transpose{\vect{1}}\mat{C}^{-1}\vect{t}_{\textnormal{cl}}
	\end{equation}
	with an expected resolution of
	\begin{equation}
		\label{eq:trackTimeResolution}
		\sigma_{\mathrm{tk}} = \left(\transpose{\vect{1}}\mat{C}^{-1}\vect{1}\right)^{-1/2}
		\,.
	\end{equation}
	\Eq{timeResolution} yields a resolution of \SI{438(16)}{\pico\second} for the track time from \eq{trackTime}, which is in agreement with the expected value from \eq{trackTimeResolution} of \SI{436}{\pico\second}. For eight uncorrelated planes with a resolution of \SI{1.02}{\nano\second}, we would expect a resolution of 
	\begin{equation*}
		\sigma = \frac{\SI{1.02}{\nano\second}}{\sqrt{8}}=\SI{361}{\pico\second}
		\,,
	\end{equation*}	
	which means that the plane correlations contribute \SI{248(9)}{\pico\second} to the track time resolution.

	\subsection{Charge based timewalk correction}
\begin{figure}[htbp]
		\centering
		\includegraphics{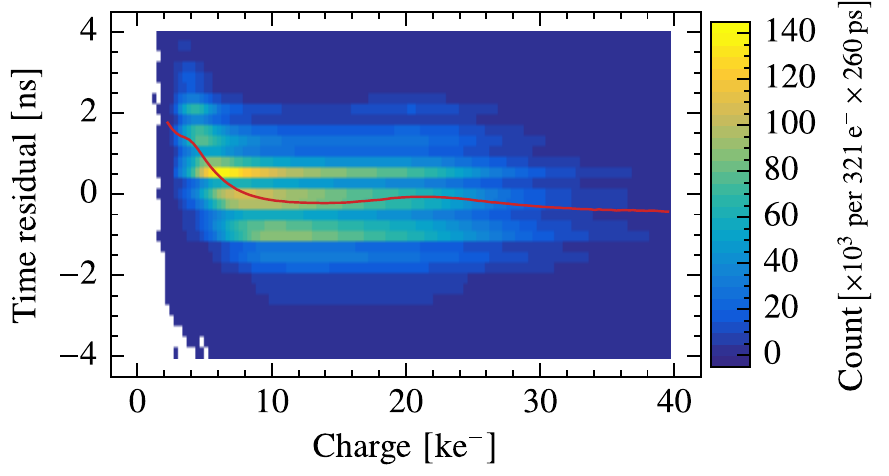}%
		\caption{Time residuals of the cluster times w.r.t. the downstream scintillator in a single telescope plane as a function of signal charge in the first cluster hit. The solid curve shows the mean time\=/residual. The horizontal banding is due to unequal bin sizes in the SPIDR TDC.}
		\label{fig:timewalkOverall}
	\end{figure}

	\Fig{timewalkOverall} shows the track time residuals with respect to the downstream scintillator as a function of signal charge. The charge bin size is chosen such that it corresponds roughly to a single \SI{25}{ns} ToT bin for high ToT values. For low charge, the mean time\=/residual behaves as can be expected for timewalk in the analog front\=/end: it decreases with increasing charge. However, at a charge of about \SI{15}{\kilo\electron} it starts to rise again, which means that timewalk is not the only effect. Using the mean time\=/residual as a correction for the cluster times nevertheless improves the track time resolution from~\SI{438(16)}{\pico\second} to~\SI{415(15)}{\pico\second}. The next section will describe a more effective method to correct for timewalk. Here we are mainly concerned with understanding the charge dependence of the time residuals, which is a prerequisite for the next section.	
	
\begin{figure}[htbp]
		\centering
		\includegraphics{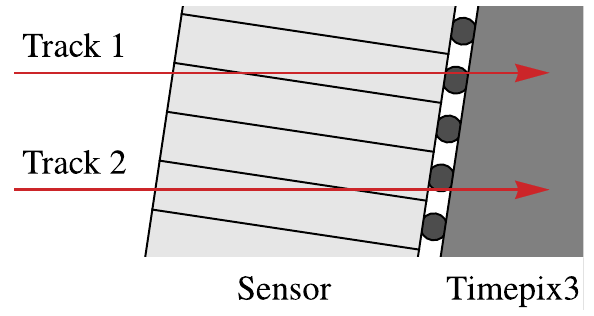}%
		\caption{Close\=/up view of two tracks in a telescope plane. The planes are angled to optimise the spatial resolution.}
		\label{fig:trackGeometryDiagram}
	\end{figure}

	Explaining why the mean time\=/residual is not a monotonically decreasing function of charge requires us to take a closer look at what happens in the sensor. For the majority of clusters, the earliest hit will be from the pixel where the track is closest to the pixel implant because this pixel has the smallest delay caused by drift time and signal induction (because the weighting field peaks near the pixels implant). We will refer to this pixel, where the track is nearest to the pixel implant, as the \textit{near pixel}. The first track in \fig{trackGeometryDiagram} is an example in which the near pixel will most likely have the earliest hit. Looking at the second track, however, we see that only a small segment of the track lies in the near pixel. Therefore, it is likely that it only receives a small portion of the total charge in the cluster, resulting in a large timewalk delay. Hence, the \textit{far pixel} in the same cluster may now have an earlier timestamp despite drift time and induction effects.
		
\begin{figure}[htbp]
		\centering
		\includegraphics{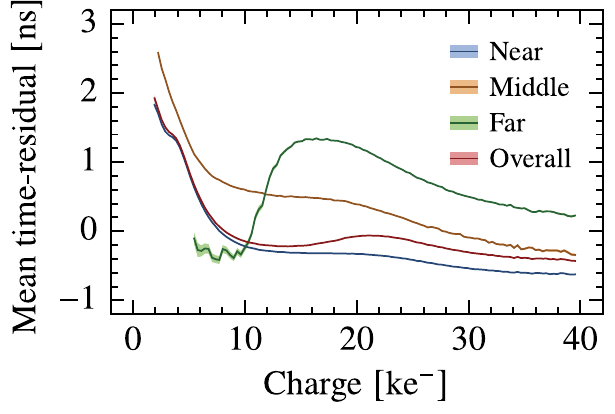}%
		\hskip10mm
		\includegraphics{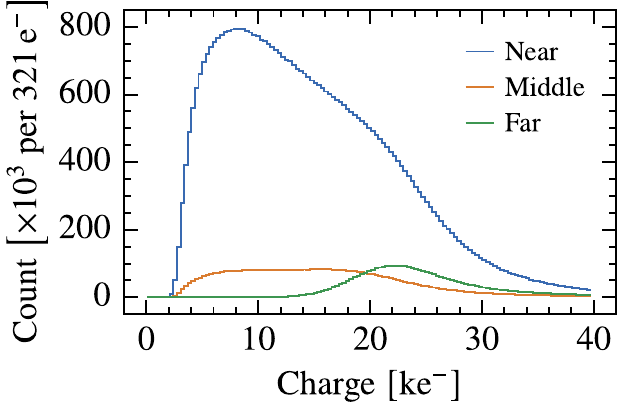}%
		\caption{Mean cluster time\=/residual as a function of charge (left) and charge distribution (right) of first hits in a cluster for a single telescope plane.}
		\label{fig:timewalkBrokenDown}
	\end{figure}

	\Fig{timewalkBrokenDown} shows the mean values of the time residuals for near and far pixels as a function of signal charge. The \textit{middle\=/pixel} curve is obtained from events where the earliest hit in the cluster is from the pixel that collected charge from the middle of the sensor. This can happen because the telescope planes are rotated around two axes, which was not shown in \fig{trackGeometryDiagram} for the sake of simplicity. All curves are monotonically decreasing functions of charge, except for the far pixels. The increasing part of the far\=/pixel events is most likely due to miscategorisation as a result of weak modes in the alignment\===a systematic offset in the plane angles can cause near\=/pixel events to be mistaken as far\=/pixel events. The local maximum in the overall mean coincides with a local maximum in the charge distribution of the far pixels. Hence, the local maximum in the mean time\=/residual can be explained by clusters that have their earliest hit in a far pixel.
	
	\subsection{Time correction using track topology and signal charge} 
	\label{sec:sensorAndTimewalkCorrections}
	In the previous section we showed that the timing behaviour as a function of signal charge depends strongly on the track topology. 
	To understand this effect we need to realise that the track topology dictates (1)~the drift time, (2)~a time offset resulting from the signal induction in a non\=/uniform weighting field \cite{Riegler:2017}, and (3)~the mean signal charge\===the track length in the sensor volume of a single pixel determines the expected charge that this pixel collects. Additionally, timewalk in the analog front\=/end also depends on signal charge, and hence also on the track topology. The result is a mixture of different effects leading to a complex dependence of the timing behaviour on track topology and signal charge fluctuations. Therefore, we construct a correction of the cluster time measurement that uses both the track intercept with the sensor as well as the signal charge in order to simultaneously compensate for drift time and signal induction effects as well as timewalk in the analog front\=/end.
	
\begin{figure}[htbp]
		\centering
		\includegraphics{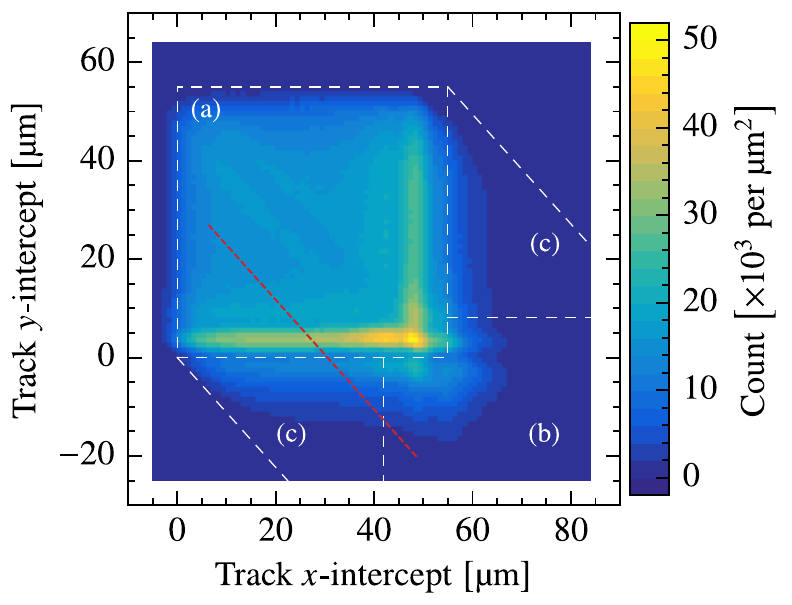}%
		\caption{Distribution of the track intercepts at the pixel implant side of the sensor. The intercept is relative to the pixel with the earliest hit in the cluster. }
		\label{fig:relTrackItsc}
	\end{figure}

	\Fig{relTrackItsc} shows the distribution of track intercepts relative to the position of the earliest pixel in the cluster. The intercepts are determined at the pixel\=/implant side of the sensor material. Region~(a) indicates the pixel of the earliest hit in the cluster and contains the near\=/pixel events, (b) contains far\=/pixel events, and (c) contains middle\=/pixel events. The red line is the projection of a single track segment onto the pixel plane. The horizontal and vertical bands that contain a higher density of tracks, as well as the diagonal band with a slightly lower density, are artefacts from the centre\=/of\=/gravity method that is used to calculate the cluster position. Some tracks lie to the left of the near\=/pixel region which is most likely due to a systematic error in the track angle originating from weak modes in the telescope alignment. This, however, does not affect the correction, since it is insensitive to an overall shift of the track intercepts.
	
\begin{figure}[htbp]
		\centering
		\includegraphics{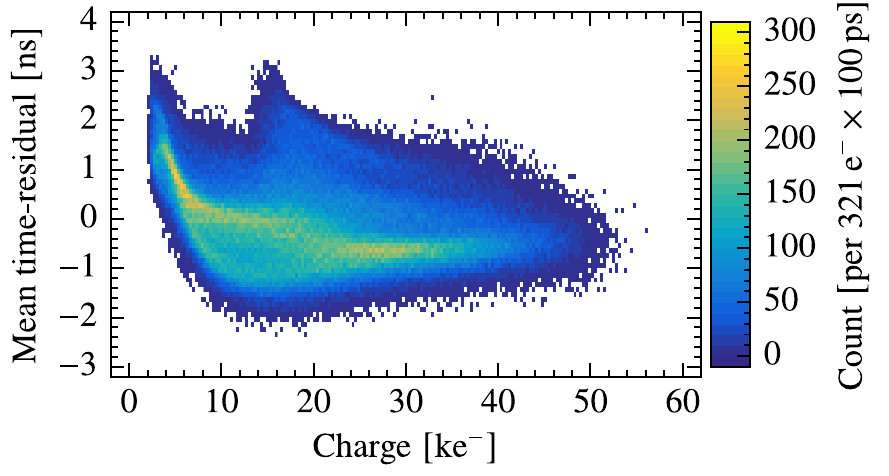}%
		\caption{Distribution of the time residual means as a function of signal charge.}
		\label{fig:correctionVsCharge}
	\end{figure}

	To calculate the time corrections for each telescope plane, we divide the events based on (1)~the relative track intercept in bins of $1\times1\;\si{\square\micro\meter}$, and (2)~the signal charge in the earliest hit into bins corresponding to roughly one ToT count ($240\text{--}\SI{320}{\electron}$ depending on the telescope plane). We create a lookup table by taking the mean time\=/residual as the correction value for each bin. \Fig{correctionVsCharge} shows how the mean values of the time residuals are distributed for all track\=/intercept bins as a function of the signal charge. For example, when the first cluster hit has a signal charge of~\SI{16}{\kilo\electron}, the mean time\=/residual can vary from \SIrange{-2}{3}{\nano\second}, depending on the location of the track intercept. Applying this correction improves the cluster and track time resolutions to about~\SI{0.83(3)}{\nano\second} and~\SI{385(14)}{\pico\second}. Hence, his method removes contributions of about~\SI{596}{\pico\second} and~\SI{209}{\pico\second} respectively. The plane correlations change only slightly, and their overall contribution to the track time resolution remains the same.
	
	\subsection{Timing systematics over the pixel matrix}
	\label{sec:pixelMatrixCorrections}
\begin{figure}[htbp]
		\centering
		\includegraphics{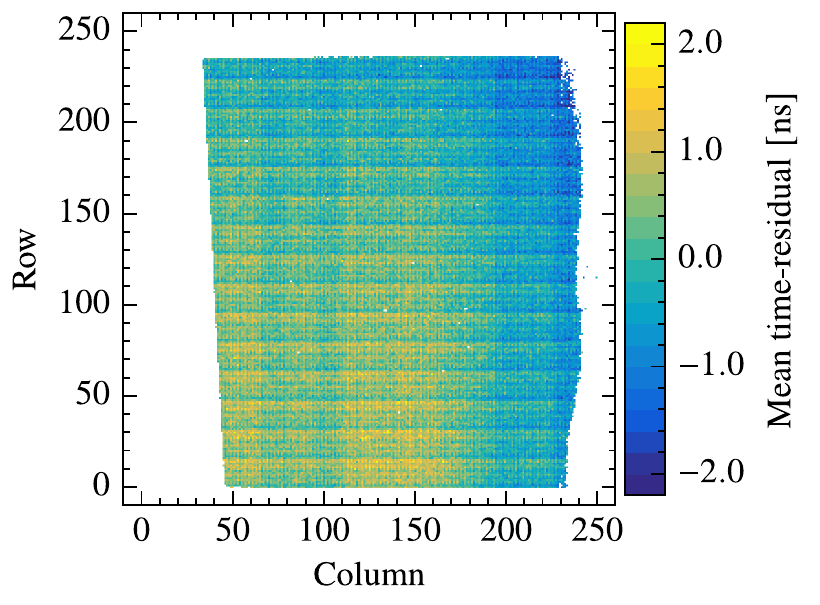}%
		\caption{Time residual means of individual pixels for a single telescope plane.}
		\label{fig:meanPixelResiduals}
	\end{figure}

	\Fig{meanPixelResiduals} shows the mean values of the time residuals for each pixel in one of the telescope planes after applying the correction from \sect{sensorAndTimewalkCorrections}. It shows that pixels can have a systematic timing error ranging from \SIrange{-2}{2}{\nano\second}. The plot also exhibits some interesting features. Firstly, there is a clear pattern visible along the columns with a 16 row periodicity. The abrupt changes that happen every 16\textsuperscript{th} row are most likely a consequence of the clock buffers that are placed along the columns for the distribution of the \SI{40}{\mega\hertz} reference clock \cite{Poikela:2015}. Secondly, odd numbered columns tend to have a slightly higher time residual than even numbered columns: on average, they are later by \SI{326(5)}{\pico\second}. Furthermore, the fine structure within blocks of 2 by 16 pixels is likely caused by a combination of fixed deviations in the frequency of the fast oscillators, and pixel-to-pixel variations in the signal propagation delay between a pixel and the fast oscillator due to differences in the capacitive loading of the traces that connect them. Lastly, there are also less regular variations: along column 150, for example, the mean time\=/residual gradually becomes smaller for larger row numbers. These gradual variations are a consequence of the power distribution over the pixel matrix.\footnote{As with the Krummenacher current (in \sect{chargeCalibration}), this issue was addressed in the second iteration of Timepix3.} 
	
\begin{figure}[htbp]
		\centering
		\includegraphics{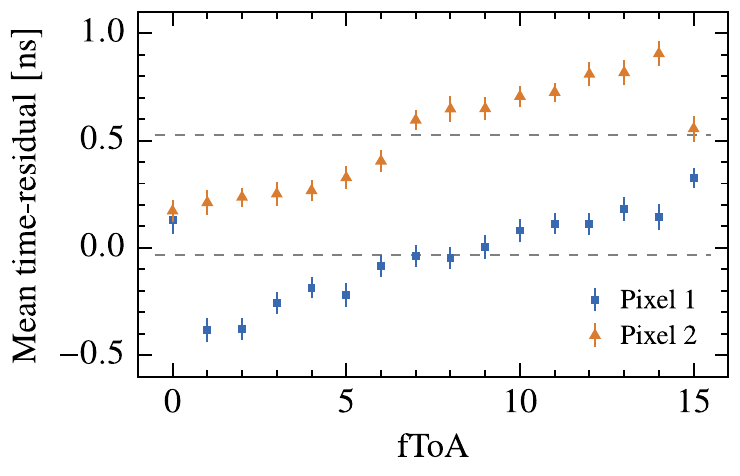}%
		\caption{Mean time\=/residual as a function of the fToA bin for two pixels in a single telescope plane. The dashed lines indicate the overall mean time\=/residual of the pixels.}
		\label{fig:meanTimeResVsFtoa}
	\end{figure}

	\Fig{meanTimeResVsFtoa} shows the mean time\=/residual for each of the 16 fToA values in two pixels. It can be seen that both fast oscillators have a frequency that is slightly too low, resulting in a constant rise of the mean time\=/residual over fToA values 1\==14. The mean time\=/residual for fToA values 0 and 15 are not necessarily in line with the others because the size of their time bins are typically different. For an fToA of 0, the time bin size is the sum of the length of the first clock cycle and the propagation delay between a pixel and the fast oscillator. For an fToA of 15, the time bin size is \SI{25}{\nano\second} minus the sum of the other time bins. The fToA can never reach a value larger than 15 because the pixel stops counting after reaching this value. We use the mean time\=/residual of each individual pixel and fToA value to correct for the systematics over the pixel matrix.
	
	\subsection{Improved timing performance}
\begin{figure}[htbp]
		\centering
		\includegraphics{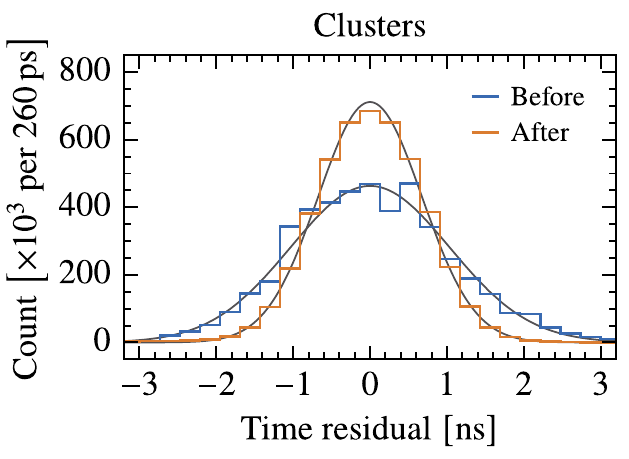}%
		\hskip10mm
		\includegraphics{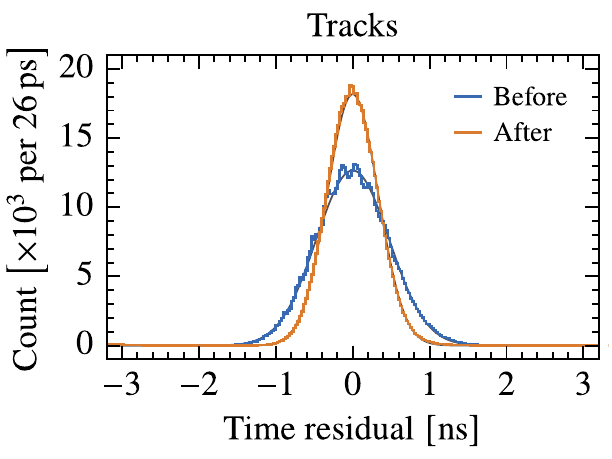}%
		\caption{Cluster and track time residuals with respect to the downstream scintillator before and after corrections. Note that the downstream scintillator resolution broadens the improved track time residual distribution by \SI{20}{\percent}.}
		\label{fig:timeResidualsBeforeAndAfter}
	\end{figure}

	After applying the charge and track\=/topology corrections from \sect{sensorAndTimewalkCorrections} as well as the pixel\=/matrix corrections from \sect{pixelMatrixCorrections}, the average cluster time resolution per plane improves from \SI{1.02(4)}{\nano\second} to \SI{650(9)}{\pico\second}. This means that we removed a contribution of \SI{0.79(5)}{\nano\second} resulting from systematic errors. The track time resolution improves from \SI{438(16)}{\pico\second} to \SI{276(4)}{\pico\second}, which means that we removed a contribution of \SI{340(21)}{\pico\second}. \Fig{timeResidualsBeforeAndAfter} shows the cluster and track time residuals with respect to the downstream scintillator before and after corrections.
	
\begin{figure}[htbp]
		\centering
		\includegraphics{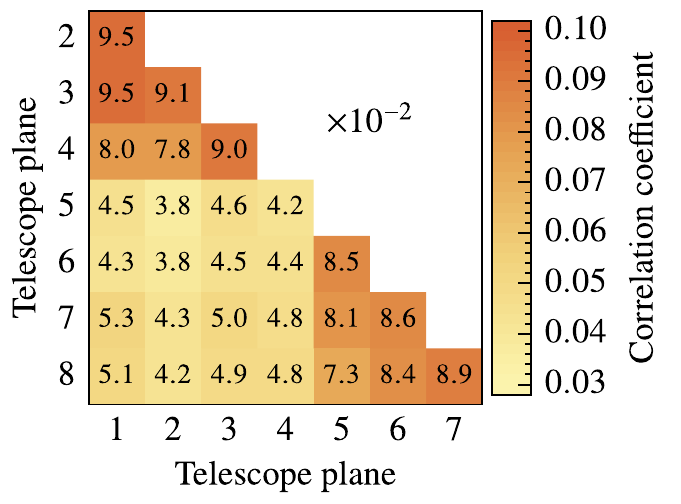}%
		\caption{Correlations between the telescope planes after corrections}
		\label{fig:correlationCoefficientsAfter}
	\end{figure}

	\Fig{correlationCoefficientsAfter} shows the correlation between planes after applying both corrections. The coefficients now range from~$0.038$ to $0.095$, showing that the planes are still correlated, though less than before. It can be seen that there is a stronger correlation between planes within the same telescope arm than between planes of two different arms. Also within the up- and downstream arms, planes that are closer together tend to have a stronger correlation. 
	
	In \sect{rawTimingPerformance}, we argued that the correlations contribute about \SI{248(9)}{\pico\second} to the track time resolution. The corrections reduced this to \SI{153(3)}{\pico\second}. This improvement is completely due to the correction of the pixel matrix systematics because it eliminates the spatial dependence of the mean time\=/residual. All planes exhibit approximately the same timing behaviour as was shown in \fig{meanPixelResiduals} and they are arranged in such a way that the row axes of all planes generally point in same direction. This results in a correlation between the planes\===when a track goes trough a region of the pixel matrix that has a large mean time\=/residual in one plane, then it is likely to go through a similar region in the other planes as well.
	
	The remaining correlations are partly caused by variations in the time offsets between the telescope planes and the scintillators. These variations happen in a continuous manner and appear to be periodic over an exact number of clock cycles from the \SI{40}{\mega\hertz} reference clock. For instance, over a period of \num{1024} clock cycles, we observe a covariance of $(\SI{134}{\pico\second})^2$ between the time offsets of planes \num{1} and \num{2} with respect to the up- and downstream scintillators, respectively. Recalling that the cluster time resolution is \SI{650}{\pico\second}, we can approximate the contribution of these covarying offsets to the correlation coefficient of the two planes as
	$\left(\SI{134}{\pico\second}/\SI{650}{\pico\second}\right)^2 \approx \SI{4}{\percent}$, which is a substantial part of the \SI{9.5}{\percent} correlation that we observe. The offsets of planes \num{1} and \num{5} exhibit a similar covariance of $(\SI{136}{\pico\second})^2$, and thus also contribute approximately \SI{4}{\percent}, which in this case is the majority of the total correlation. The fact that we observe synchronous variations in the offsets of planes that are connected to different SPIDR modules and with respect to different scintillators seems to suggest that they originate from the reference clock.\footnote{This is only possible because there is a time offset of about \SI{250}{\nano\second} between the telescope planes and the scintillators (due to electronics and cabling) which causes their measurements to be referenced to different clock edges.} This view is substantiated further by matching peaks in the phase noise spectrum of the clock. Regardless of the exact cause, we estimate that correcting for these effects could improve the track time resolution to about \SI{240}{\pico\second}, mainly by further reduction of the correlations.

	We have excluded any further spatially induced correlations by looking at tracks that go through a small fiducial area. In a similar way, we also found no evidence of a time correlation induced by a correlation in track angle. Time of flight also does not cause correlations since this would only change the time offsets between the planes, which are removed by the time alignment. Furthermore, significant event\=/to\=/event variations in the time of flight are unlikely.
	
\begin{figure}[htbp]
		\centering
		\includegraphics{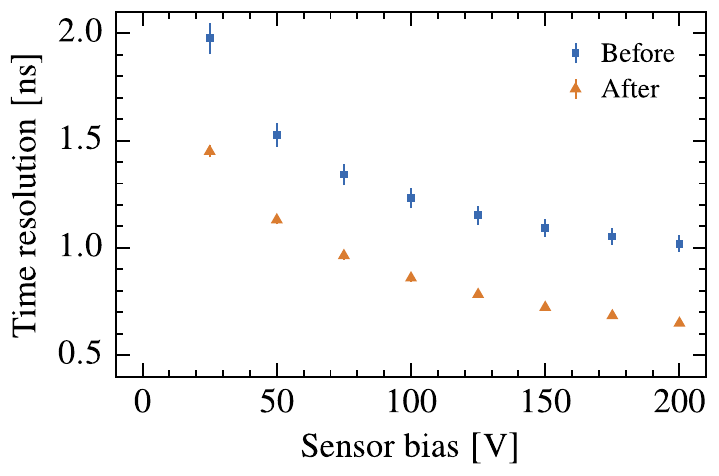}%
		\caption{Overall cluster time resolution as a function of sensor bias voltage before and after corrections.}
		\label{fig:biasScan}
	\end{figure}

	Subtracting the TDC resolution of \SI{451}{\pico\second} from the improved cluster time resolution, shows that the sensor and analog front\=/end contribute about \SI{467}{\pico\second}, which is comparable to the TDC resolution. \Fig{biasScan} shows a sensor bias\=/voltage scan of the telescope. It can be seen that the cluster time resolution is still improving at a sensor bias voltage of \SI{200}{\volt}. However, we did not increase the bias voltage beyond this point due to the risk of breakdown as already mentioned above.
	
	\section{Conclusion and outlook}
	We performed a detailed study of the timing performance of the LHCb VELO Timepix3 Telescope, which resulted in a thorough understanding of time measurements with Timepix3. Using an independent scintillator system, we measured the time resolution before any corrections to be \SI{1.02(4)}{\nano\second} for single clusters and \SI{438(16)}{\pico\second} for tracks consisting of eight clusters. Correlations between the cluster time measurements of different planes result in a track time resolution that is worse than what can be expected for eight uncorrelated planes: $\SI{1.02(4)}{\nano\second}/\sqrt{8}=\SI{358(13)}{\pico\second}$. As discussed in \sect{pixelMatrixCorrections}, most of these correlations are due to a spatial dependence of the timing behaviour in the planes. 
	
	In \sect{sensorAndTimewalkCorrections} we presented a method that uses the track topology in combination with the signal charge to simultaneously correct for timewalk and sensor effects. This method could be extended to work with variable track directions by first binning on two angles that define the track direction in the sensor, and subsequently binning on a track intercept. In a 4D tracker the time measurements are in principle necessary for the track reconstruction itself. This means that in the worst case scenario an iterative method might be required to apply this correction. However, if the time measurement is only essential in finding the correct primary vertex\===the hadronic collision point\===then the track reconstruction will be relatively insensitive to the specific vertex since the candidates, out of which the 4D tracking algorithm has to select the correct one, are spatially close together since they belong to the same bunch crossing. As a result, the reconstructed track before finding the correct primary vertex is already precise enough to determine the cluster time corrections, and hence the method can be applied directly without iteration.

	Correcting for systematic errors from the pixel matrix also significantly improves the timing performance of the individual planes. Additionally, it removes the spatially induced correlations between the planes, which further improves the track time resolution.

	After applying corrections, the overall cluster time resolution improves from \SI{1.02(4)}{\nano\second} to \SI{650(9)}{\pico\second}. This implies that we have removed a contribution of \SI{0.79(5)}{\nano\second} resulting from systematic errors. The track time resolution improves from \SI{438(16)}{\pico\second} to \SI{276(4)}{\pico\second}. This is an improvement of \SI{340(21)}{\pico\second} of which we can attribute \SI{278(24)}{\pico\second} to an improvement in the cluster time resolution and \SI{195(12)}{\pico\second} to removing part of the plane correlations.

	\acknowledgments
	We express our gratitude to Wiktor Byczynski and Raphael Dumps at CERN for their vital support during the test beam period. We also thank our colleagues in the CERN accelerator departments for the excellent performance of the SPS. This research was funded by the Dutch Research Council~(NWO).
	
	\bibliography{bibliography}
	
\end{document}